\documentclass{article}
\usepackage{amsmath}
\usepackage{graphicx}
\usepackage{amsfonts}
\usepackage{amssymb}
\newtheorem{theorem}{Theorem}

\newtheorem{corollary}[theorem]{Corollary}

\newtheorem{definition}[theorem]{Definition}
\newtheorem{example}[theorem]{Example}

\newtheorem{proposition}[theorem]{Proposition}

\begin{document}

\title{A Quantum Information Theoretical Model for Quantum Secret Sharing Schemes}
\author{Hideki Imai${}^{\dagger}$\qquad{} J\"orn M\"uller--Quade\thanks{Universit\"at
Karlsruhe, Fakult\"at f\"ur Informatik, IAKS Beth, Postfach 6980, 76128
Karlsruhe, Germany. Email: \texttt{muellerq@ira.uka.de}}\\{} Anderson C. A. Nascimento\footnote{Imai Laboratory, Information and
Systems, Institute of Industrial Science, University of Tokyo, 4--6--1 Komaba,
Meguro--ku, Tokyo 153--8505, Japan. Email:
\texttt{anderson@imailab.iis.u-tokyo.ac.jp}, \texttt{imai@iis.u-tokyo.ac.jp}
}\qquad{} Pim Tuyls\footnote{Philips Research, Mailbox WY7.12, Prof.~Holstlaan
4, 5656 AA Eindhoven, Netherlands. Email: \texttt{pim.tuyls@philips.com}
}\qquad{} Andreas Winter\footnote{School of Mathematics, University of
Bristol, University Walk, Bristol BS8 1TW, United Kingdom. Email:
\texttt{a.j.winter@bris.ac.uk}}\qquad{}}
\date{14 October 2003}
\maketitle

\begin{abstract}
In this paper we introduce a quantum information theoretical model for quantum
secret sharing schemes. We show that quantum information theory provides a
unifying framework for the study of these schemes. We prove that the
information theoretical requirements for a class of quantum secret sharing
schemes reduce to only one requirement (the recoverability condition) as a
consequence of the no-cloning principle. We give
also a shorter proof of the fact that the size of the shares in a quantum
secret sharing scheme must be at least as large as the secret itself.
\end{abstract}

\section{Introduction\label{Section1}}

Quantum secret sharing has been an active area of research in quantum
information theory \cite{CGL}\cite{HBB}\cite{TGZ}\cite{Gott}. In a quantum
secret sharing protocol, a dealer shares an unknown quantum state with a set
of players such that authorized subgroups of players can recover the quantum
state, but unauthorized subgroups cannot get any information on it. Quantum
secret sharing was first introduced in \cite{HBB}, where Hillery \emph{et al.}
proposed a scheme to share a single qubit between two players. In \cite{CGL}
Cleve, Gottesman and Lo presented a more general scheme where a dealer can
share an unknown quantum state with a set of players in a way that only groups
with more than a given number of players, $t$, can recover the original secret
and collusions of players with less than $t$ players have no information about
it. Given that the total number of players is $n$, this is a quantum
$(t,n)$-threshold scheme. The construction in \cite{CGL} was based on quantum
error correcting codes. Constructions for general access structures were
presented in \cite{Gott} by Gottesman and in \cite{Smith} by Smith.

In classical secret sharing, and in classical cryptography in general,
information theory has played a major role when designing and evaluating
cryptographic primitives and protocols 
\cite{Maurer}\cite{Shannon}\cite{CSS}\cite{CSS2}. 
It is a natural question to investigate the properties of their
quantum mechanical counterparts. In this contribution, we introduce a quantum
information theoretical model for quantum secret sharing schemes. We show that
quantum information theory provides a unifying framework for the study of
these schemes. Additionally, we prove that the information theoretical
requirements for some quantum secret sharing schemes differ from the ones for
their classical counterparts. Moreover, we give a shorter proof of the fact
that the size of the shares in a quantum secret sharing scheme must be at
least as large as the secret itself. This result was first stated in
\cite{Gott}.

The paper is organized as follows: in Section 2, we review some important
concepts of quantum information theory that are used in this paper. Section 3
introduces our model for quantum secret sharing schemes. We show that the
recoverability requirement for pure state quantum secret sharing protocols
implies the secrecy one in Section 4. In Section 5, we present a new and
shorter proof of the fact that the sizes of the shares of a quantum secret
sharing scheme are at least as large as the size of the quantum secret being
shared. Finally, we conclude in Section 6.

\section{Preliminaries}

\subsection{Quantum Information Theory}

We briefly review some important concepts of quantum information theory that
will be used through the paper. For a nice introduction to the subject we
suggest the references \cite{Preskill} and \cite{NC}. We consider finite
dimensional quantum systems with $m$ degrees of freedom which are modeled by
the algebra of $m\times m$ matrices over the complex numbers, here denoted by
$\mathcal{M}_{m}$. The state of a system $X$ is described by its density
matrix $\rho_{X}\in\mathcal{M}_{m}$.

The quantum entropy of a quantum system $X$ with a density matrix $\rho_{X}
\in\mathcal{M}_{m}$ is defined as in~\cite{NC}

\[
\mathsf{S}(X)=-\text{Tr}(\rho_{X}\log\rho_{X})=-\underset{1\leq j\leq m}{\sum
}\lambda_{j}\log\lambda_{j},
\]
where $\lambda_{1},\ldots,\lambda_{m}$ are the eigenvalues of $\rho_{X}$. The
quantum entropy can be interpreted as the average number of qubits necessary
to describe a realization of the system $X$ \cite{Schuma}.

Quantum entropies generalize the classical Shannon entropies. For a random
variable $A$ taking values in an alphabet $\mathcal{A}=\{a_{1},\ldots,a_{n}\}$
one has

\[
\mathsf{H}(A)=-\underset{a\in\mathcal{A}}{\sum}p(a)\log p(a),
\]
where the random variable $A$ takes the value $x$ with probability $p(x)$.
When all the quantum states that compose the quantum mixture $\rho_{X}$ are
orthogonal, $\mathsf{S}(X)$ reduces to $\mathsf{H}(X).$

Conditional entropy is a very important tool used to analyze classical
systems. For two random variables $A$ and $B$ taking values in the alphabets
$\mathcal{A}$ and $\mathcal{B}$ respectively, the conditional entropy is
defined as:

\[
\mathsf{H}(A|B)=\underset{a\in\mathcal{A},b\in\mathcal{B}}{\sum}p(a,b)\log
p(a|b).
\]
In order to analyze quantum secret sharing schemes precisely, it is important
to generalize classical conditional entropies to the quantum domain.

Let $XY$ be a bipartite quantum system represented by a density matrix
$\rho_{XY}$ living on the Hilbert space $\mathcal{H}_{XY}=\mathcal{H}
_{X}\otimes\mathcal{H}_{Y}.$ \ The subsystems $X$ and $Y$ will be represented
by the partial traces $\rho_{X}=\text{Tr}_{Y}\rho_{XY}$ and $\rho
_{Y}=\text{Tr}_{X}\rho_{XY}.$ The quantum entropy of a quantum system $X$
conditional on another quantum system $Y$ can be defined as
(see \cite{NC}):

\begin{equation}
\mathsf{S}(X|Y)=\mathsf{S}(XY)-\mathsf{S}(Y),
\end{equation}
where $\mathsf{S}(XY)=-\text{Tr}(\rho_{XY}\log\rho_{XY})$ and $\mathsf{S}
(Y)=-\text{Tr}(\rho_{Y}\log\rho_{Y})$.

One can understand the quantum conditional entropy as the ignorance about the
quantum system $X$ when having full knowledge of $Y.$

It is important to stress that it is possible to define different versions of
quantum conditional probabilities. However, a nice point about the definition
used here is that several well known properties of classical conditional
entropies are valid in the new scenario~\cite{AC1}\cite{AC2}\cite{Bell}\cite{Sep}.

In spite of these similarities in the formulae, quantum conditional entropies
are qualitatively different from their classical counterparts. For example,
quantum conditional entropies can be negative while classical conditional
entropies are always non-negative. It means that in quantum systems,
sometimes, the entropy of the entire quantum system can be smaller than the
entropy of one of its subsystems. This is the case for the so called entangled
systems. Another consequence of the negativity of quantum conditional
entropies is that proofs from classical information theory do not usually
straightforwardly apply to the quantum scenario, since they often rely on
the non-negativity of conditional entropies.

Similarly, a quantum counterpart of the classical mutual information is
defined (see~\cite{NC}) as follows:

\[
I(X:Y)=\mathsf{S}(X)+\mathsf{S}(Y)-\mathsf{S}(XY)\geq0.
\]
It should be remarked that the quantum information does not only measure
quantum correlations between two systems. It includes both quantum and
classical correlations \cite{AC1}.

The quantum mutual information can be interpreted as the information on the
quantum state $X$ that is conveyed by $Y$. Indeed, it is $0$ iff
the state of XY is a product: $\rho_{XY}=\rho_X\otimes\rho_Y$.

The subadditivity, strong subadditivity and the Araki-Lieb inequalities of
quantum entropies will be heavily used when deriving our results. For the
convenience of the reader we state these results here \cite{NC}.

The subadditivity of quantum entropies tells us that for a composite quantum
system $XY$, the following inequality holds:

\[
\mathsf{S}(XY)\leq\mathsf{S}(X)+\mathsf{S}(Y).
\]
The Araki-Lieb inequality is stated as:

\[
\mathsf{S}(XY)\geq|\mathsf{S}(X)-\mathsf{S}(Y)|
\]
where $|\cdot|$ denotes the absolute value.

Finally, the strong subadditivity states that for any composite quantum system
$XYZ$, the following inequality holds:

\[
\mathsf{S}(XYZ)+\mathsf{S}(Y)\leq\mathsf{S}(XY)+\mathsf{S}(YZ).
\]
It is easy to show that the following inequality is a consequence of the
strong subadditivity of quantum entropies:

\[
I(X:Y)\leq I(X:YZ)
\]
for any tripartite system $XYZ.$

\subsection{Classical Secret Sharing Schemes}

As stated in Section \ref{Section1}, a secret sharing scheme is a
protocol that enables a dealer $\mathcal{D}$ to share a secret $S_{i}$
from a set of $n$ possible secrets $\Omega=\{S_{1},\ldots,S_{n}\}$
with a set of players $\mathcal{P}$ so that the members of an
authorized group are able to recover $S_{i}$, but no other members can
get any information about the secret $S_{i}$. The authorized groups
will be defined by an access structure $\Gamma
\subset2^{\mathcal{P}}$, a family where each element is an authorized
group.  More precisely, for a set of participants
$\mathcal{P}=\{P_{1},\ldots,P_{m}\}$ and a dealer $\mathcal{D}$, the
access structure $\Gamma\subset2^{\mathcal{P} }$ is a family of
subsets of $\mathcal{P}$ containing the sets of participants qualified
to recover the secret. Monotonicity is a natural requirement of an
access structure, i.e. if $X\in\Gamma$ and $X\subset X^{\prime}$ then
$X^{\prime}\in\Gamma$. There is one operation $\Lambda$ which chooses
randomly with a given distribution a tuple of shares $\in
\Omega_1\times\dots\times\Omega_m$ for a given secret $\in \Omega$. By
$\Lambda_j: \Omega\rightarrow\Omega_j$ we denote the restriction of
the operation $\Lambda$ to one player $P_j$ determining its share.  To
obtain consistent shares all $\Lambda_j$, $j\in {\cal P}$ refer to the
same execution of the operation $\Lambda$.  Assuming a probability
distribution on $\Omega$, the secret $S$ and the shares $\Lambda_j$
become random variables.  A secret sharing scheme is called
\textit{perfect} if:

\begin{enumerate}
\item  Any set of qualified participants $X\in\Gamma$ can uniquely determine
the secret $S$, i.e. $\mathsf{H}(S|\Lambda_j:j\in X)=0$ .

\item  None of the subsets $X\subset\mathcal{P}$, $X\notin\Gamma$ can get 
information about the secret $S$, i.e. $\mathsf{H}(S|\Lambda_j:j\in
X)=\mathsf{H}(S)$.
\end{enumerate}

When $|\mathcal{P}|=m$ and $\Gamma=\{B\subseteq\mathcal{P}:|B|\geq t\}$ the
secret sharing scheme is called a $(t,m)-$threshold scheme.

\section{A Model for Quantum Secret Sharing Schemes}

In this Section, we provide a formal definition of a secret sharing scheme
based on quantum entropies. In a quantum secret sharing protocol, a dealer $D$
wants to share a quantum state $|X\rangle$ with a set of players $\mathcal{P}$
according to a given access structure $\Gamma\subset2^{\mathcal{P}}$. The
access structure $\Gamma$ is a family that lists all the subsets of players
that can recover the quantum secret $|X\rangle.$

In our model, the quantum secret $|X\rangle$ is chosen from a set of possible
quantum secrets $\mathcal{X}=\{$ $|X_{1}\rangle,|X_{2}\rangle,\ldots
,|X_{n}\rangle\}.$ \ The \textit{a priori} probability that the secret
$|X_{i}\rangle$ is chosen is $p_{i}.$ The quantum secret $S$ can thus be
represented by the quantum mixture:

\[
\rho_{S}=p_{1}|X_{1}\rangle\langle X_{1}|+p_{2}|X_{2}\rangle\langle
X_{2}|+\ldots+p_{n}|X_{n}\rangle\langle X_{n}|.
\]
We assume the states $|X_{i}\rangle$ to be pure states. The set of possible
quantum shares given to a player $P\in\mathcal{P}$ and any quantum state that
he may possess are, for simplicity of notation, also denoted by $P$. Its
density matrix is represented by $\rho_{P}$.

Each quantum secret is assumed to lie in a $n$-dimensional Hilbert space
$\mathcal{H}_{S}$. We will model the shares of the players by quantum systems.
The Hilbert space in which the share of player $i$ lives is denoted by
$\mathcal{H}_{i}$. If $A$ is a subset of $\mathcal{P}$, then we will denote
the Hilbert space that describes the shares of players in $A$ by
$\mathcal{H}_{A}=\otimes_{a\in A}\mathcal{H}_{a}$.
Let us introduce a reference system $R$ with Hilbert space
$\mathcal{H}_{R}$ and a purification, denoted
$|SR\rangle\in\mathcal{H}_S\otimes\mathcal{H}_R$: i.e., after tracing out
$R$, one recovers $\rho_S$ \cite{NC}.
A distribution of shares is given by a completely positive map
\begin{equation}
  \Lambda_{D}:S(\mathcal{H}_{S})\rightarrow S(\mathcal{H}_{1}\otimes
     \ldots\otimes\mathcal{H}_{m})
\end{equation}
where $S(\mathcal{H}_{A})$ represents the state space of the system $A$.
Note that by the Stinespring dilation theorem we can always make $\Lambda_D$
an isometry (as we shall henceforth assume to be the case),
by adding a player $P_{m+1}$
\cite{NC} (and trivially extending the access structure).
We denote by $|RP_{1}...P_{m}\rangle$ the state of the quantum system
$RP_{1}...P_{m}$ after applying $\Lambda_D$ to $S$
(and the identity to $R$). We denote by $A$, both a given subset of players
$A=\{P_{1},\ldots,P_{j}\}\subseteq$ $\mathcal{P}$ and the quantum shares which
are in the possession of those respective players. The set of all the quantum
shares which are distributed to the players in $\mathcal{P}$ is denoted by $Y$.

\begin{definition}
Let $R$ be a reference system such that $SR$ is in a pure state. A quantum
secret sharing protocol realizing an access structure $\Gamma$ is a complete
positive map which generates quantum shares $\{P_{1},\ldots,P_{m}\}$ from a
quantum secret $\rho_{S}=p_{1}|X_{1}\rangle\langle X_{1}|+p_{2}|X_{2}
\rangle\langle X_{2}|+\ldots+p_{n}|X_{n}\rangle\langle X_{n}|$, and
distributes these shares among a set of players $\mathcal{P}$,
$\vert$
$\mathcal{P}$
$\vert$
$=m$ such that:

\begin{enumerate}
\item  For all $A\in\Gamma$ we have that $I(R:A)=I(R:S)$, or equivalently, as
proved in \cite{SN}, for $A\in\Gamma$ there exists a completely positive map
$T_{A}:\mathcal{H}_{A}\rightarrow\mathcal{H}_{S}$ such that
\begin{equation}\begin{split}
  {\rm id}_{R}\otimes T_{A}: &S(\mathcal{H}_{R}\otimes\mathcal{H}_{A})\rightarrow
                                S(\mathcal{R}\otimes\mathcal{H}_{S})              \\
                             &\rho_{RA}\mapsto|RS\rangle
\end{split}\end{equation}

\item  For all $A\notin\Gamma$ we have that $I(R:A)=0.$
\end{enumerate}
\end{definition}

The requirement $1$ means that the entanglement between the reference state
and the secret is preserved when it is recovered by authorized players
(\emph{recoverability requirement}). Actually, we remark that the equality
$I(R:A)=I(R:S)$ is equivalent to saying that the coherent information
$I_{e}=\mathsf{S}(A)-\mathsf{S}(RA)$ equals the entropy of the secret
$\mathsf{S}(S).$ In \cite{SN}, it was proven that this condition is necessary
and sufficient for quantum error correction. In our case, this means that an
authorized group can reconstruct the secret exactly. Consequently, our
requirement $1$ implies a relation between quantum error correcting codes and
quantum secret sharing schemes. More in particular, this means that one can
see the mapping $\Lambda_{D}$ followed by restricting to the systems $A$
as a quantum noisy channel and the recovery process as quantum error correction.

The requirement $2$ means that unauthorized groups cannot recover any
state which is correlated to $R$, and consequently with $S$
(\emph{secrecy requirement}). Note that monotonicity is also naturally
embedded in this definition. We remind the reader that the second
requirement means that the state of the system $AR$ is a product
state, hence $A$ and $R$ are independent.  Consequently, the relative
entropy $\mathsf{S}(S|A)$ is equal to $\mathsf{S}(S).$

In the next Section we prove that, in contrast to classical schemes, the
recoverability requirement implies the secrecy requirement in some quantum
secret sharing schemes.

\begin{example}
We briefly illustrate our definition for the case of the $(2,3)$ threshold
secret sharing scheme mentioned in~\cite{CGL}. For sake of clarity, we repeat
the scheme here. The secret is an arbitrary three dimensional quantum state,
i.e. $\rho_{S}=1/3\sum_{i=0}^{2}|i\rangle\langle i|$. The encoding scheme that
encodes the shares for the different players is given by an isometry
$U_{S}:\mathbb{C}^{3}\rightarrow\mathbb{C}^{3}\otimes\mathbb{C}^{3}\otimes\mathbb{C}^{3}$
mapping
\begin{align*}
U_{S}:\alpha|0\rangle+\beta|1\rangle+\gamma|2\rangle\mapsto
        &\phantom{+}\alpha(|000\rangle+|111\rangle+|222\rangle)\\
        &          +\beta(|012\rangle+|120\rangle+|201\rangle)\\
        &          +\gamma(|021\rangle+|102\rangle+|210\rangle)
\end{align*}
We note that the operator $U_{S}$ induces a completely positive map
$\Lambda_{D}$ on $\mathcal{M}_{3}$. As $S$ is a completely mixed state, its
purification on $\mathbb{C}_{3}\otimes\mathbb{C}_{3}$ (entanglement with the
reference system) looks as follows $|RS\rangle=1/{\sqrt{3}}\sum_{i=0}^{2}|i\rangle
\otimes|i\rangle$. The system $RA$ (for $A=\mathcal{P}$) can then be described as follows,
\begin{align*}
|RA\rangle &=(\mathbf{1} \otimes U_{S})|RS\rangle\\
           &=\frac{1}{3}(|0000\rangle+|0111\rangle+|0222\rangle+|1012\rangle+|1120\rangle\\
           &\phantom{===}+|1201\rangle+|2021\rangle+|2102\rangle+|2210\rangle)
\end{align*}
It follows immediately that $I(R:S)=2\log3$. When we take $A={1,2}$, i.e. $A$ is an
authorized set, it readily follows from the previous equation that
$\mathsf{S}(A)=2\log3$ and $\mathsf{S}(RA)=\log3$. Hence, as $I(R:A)=2\log3$,
the recoverability requirement is satisfied. If on the other hand $A={1}$, i.e
$A$ is not authorized, then it follows that the system $RA$ is in the product
state $\frac{1}{3}\mathbf{1}\otimes\frac{1}{3}\mathbf{1}$. Hence the secrecy
condition is satisfied.
\end{example}

We note that the quantum entropy of the mixed state
$\rho_{S}$ can be understood as the minimum number of qubits necessary to
faithfully store the quantum secret (see \cite{Schuma}). The
same applies for the quantum mixtures representing the quantum shares of
the players.

A fundamental issue when dealing with secret sharing schemes is the amount of
data that must be given to the set of players. The smaller the amount of data
given to the set of players the better. This issue becomes even more
important when dealing with quantum secret sharing. As quantum data is
expensive and hard to deal with, it would be desirable to use as little
quantum data as possible in order to share an unknown quantum state.
Therefore, the analysis of the size of shares in quantum secret sharing
schemes is an important research subject \cite{And}. Based on the model
introduced earlier and on classical equivalents~\cite{CSS}, we define two
important quantities related to the size of the shares in a quantum secret
sharing scheme.

\begin{definition}
The \emph{quantum information rate} of a secret sharing scheme which shares a
quantum secret state $S$ with a set of players $\mathcal{P}$ realizing an
access structure $\Gamma$ is given by the following expression: $r=\frac
{\mathsf{S}(S)}{\max_{X\in\mathcal{P}}\mathsf{S}(X)}.$

The \emph{average quantum information rate} of a secret sharing scheme which
shares a quantum secret state $S$ with a set of players $\mathcal{P}$
realizing an access structure $\Gamma$ is given by the following expression:
$\bar{r}=\frac{\mathsf{S}(S)|\mathcal{P|}}{\sum_{X\in\mathcal{P}}
\mathsf{S}(X)}$.
\end{definition}

We shall demonstrate how quantum information theoretical tools can be
used to prove lower bounds on the size of the quantum shares in a quantum
secret sharing scheme.

\section{Relation between the Recoverability and Secrecy Requirement}

\label{sec-relation}

In this section, we prove that the recoverability requirement as stated in the
last section, implies the secrecy requirement for some quantum secret sharing schemes.
This result means that, for some access structures, if an authorized set of players is able to recover a
quantum secret at all, the unauthorized players have no information about the
secret. This is a consequence of the no-cloning theorem and stands in sharp
contrast with classical secret sharing schemes.

We first introduce the notion of coexistence. We say that a set of shares $A$
coexists with a secret $S$ if there exists a completely positive map $T$ from
$A^{\prime}=\mathcal{P}\setminus A$ to the system $S$ such that the secret $S$
can be recovered. More precisely, this is given by a completely positive map
$T$ such that
\begin{equation}
\text{id}_{R}\otimes T:|RA^{\prime}\rangle\longmapsto|RS\rangle.
\end{equation}
In the following proposition (which was
first observed by Gottesman \cite{Gott}), we prove that if a quantum state $A$ can coexist
with the quantum secret $S$, then $A$ should have no correlation with $S$,
that is $A$ cannot be used to recover $S$.

\begin{proposition}
Given a quantum secret $\rho_{S}=p_{1}|X_{1}\rangle\langle X_{1}|+p_{2}
|X_{2}\rangle\langle X_{2}|+\ldots+p_{n}|X_{n}\rangle\langle X_{n}|$ and a
reference system $R$ such that $RS$ is in a pure state. If $A$ can coexist with
the quantum secret $S$, then $I(R:A)=0,$ i.e. $RA$ is in a product state.
\end{proposition}

\noindent\textbf{Proof. }\noindent From the strong subadditivity property of
quantum entropies, it follows that
\[
I(A:R)\leq I(A:SR).
\]
On the other hand we have
\begin{align*}
I(A:SR)  &  =\mathsf{S}(A)+\mathsf{S}(SR)-\mathsf{S}(SRA)\\
&  \leq\mathsf{S}(A)+\mathsf{S}(RS)-\mathsf{S}(A)+\mathsf{S}(RS).
\end{align*}
The last inequality follows from the Araki-Lieb inequality. Since $RS$ is in a
pure state it follows that: $\mathsf{S}(RS)=0$, and hence that
$I(A:R)\leq0$. Since the mutual quantum information is non-negative the
proposition follows.

\hfill$\blacksquare$

\begin{theorem}
For quantum secret sharing schemes where unauthorized sets of players are the complement of authorized sets the recoverability requirement implies the
secrecy requirement.
\end{theorem}

\noindent\textbf{Proof. }In quantum secret sharing schemes where
unauthorized sets of players are the complement of authorized sets, all the
quantum states in possession of unauthorized sets of players coexist with the
secret, since it can be recovered by the authorized players. Therefore, from
Proposition 4, we know that there is no correlation between the quantum states
of unauthorized sets of players and the secret.

\hfill$\blacksquare$

\section{A Lower Bound on the Size of the Shares}

In this section we give a proof that the size of the shares in a quantum
secret sharing scheme must be as large as the size of the secret itself. This
theorem was first proved in~\cite{Gott}. However, our proof is based on quantum
entropies and it is simpler than the original one: it follows from the
subadditivity property of quantum entropies~\cite{Preskill}\cite{NC}.

\begin{theorem}
In any quantum secret sharing scheme realizing an access structure $\Gamma$
for any subsets of players $A$ and $B$ such that $A,B\notin\Gamma$ but $A\cup
B\in\Gamma$ it holds that $\mathsf{S}(A)\geq\mathsf{S}(S)$ where $S$ is the
secret being shared.
\end{theorem}

\noindent\textbf{Proof. }From the fact that $A\cup B\in\Gamma$ and Def.~1 we have that
\[
\mathsf{S}(AB)-\mathsf{S}(RAB)=\mathsf{S}(R).
\]
Applying the Araki-Lieb inequality to $\mathsf{S}(RAB)$, we get
\[
\mathsf{S}(AB)-\mathsf{S}(RA)+\mathsf{S}(B)\geq\mathsf{S}(R).
\]
Since $I(A:R)=0,$ it follows that $\mathsf{S}(RA)=\mathsf{S}(A)+\mathsf{S}
(R)$, and this together with the last inequality gives us:

\[
\mathsf{S}(AB)-\mathsf{S}(A)+\mathsf{S}(B)\geq2\mathsf{S}(R).
\]
Using the subadditivity property and the fact that $\mathsf{S}(R)=\mathsf{S}
(S),$ it follows that

\[
\mathsf{S}(B)\geq\mathsf{S}(S)
\]
which proves the theorem.

\hfill$\blacksquare$

From Definition 3 and Theorem 6, Corollary 7 follows:

\begin{corollary}
The quantum information rate and the average quantum information rate of a
secret sharing scheme are lower bounded by $1$.
\end{corollary}

\hfill$\blacksquare$

Another concept closely related to quantum secret sharing is the quantum
Vernam cipher, introduced by Leung in \cite{IBM}. In a quantum Vernam cipher,
a sender, Alice, wants to send a quantum state to a recipient, Bob, such that
an eavesdropper, Eve, when intercepting this secret quantum state has no way
to get any knowledge on it. To perform so, they share in advance entangled
states and use it as a quantum key. So, in a quantum Vernam cipher, the key
and the message are quantum states. Actually, the quantum Vernam cipher is an
implementation of a quantum secret sharing scheme. If Alice's part of the key is
represented by the quantum system $A$, Bob's by the quantum system
$B$, the encrypted message is represented by the quantum system $M$, and
the quantum cleartext resides in the secret system $S$, the quantum Vernam
cipher can be described by the following access structure $\Gamma
=\{(A,M),(B,M)\}$. Therefore, the following corollary holds.

\begin{corollary}
In a quantum Vernam cipher the size of the key is as large as the size of the
message to be transmitted.
\end{corollary}

\hfill$\blacksquare$

We remark that although the dimension of the quantum keys in a quantum Vernam
cipher is as large as the dimension of the quantum message, the quantum scheme
presents an advantage over its classical counterpart: the quantum keys can be
recycled \cite{IBM}.

\section{\bigskip Conclusions}

We introduced a quantum information theoretical model for quantum secret
sharing schemes. This model provided new insights into the theory of quantum
secret sharing. We proved that the recoverability requirement implies the
secrecy requirement for a class of quantum secret sharing schemes by
giving an information theoretical argument for Gottesman's result
that the complement of an authorized set is unauthorized. Also, we
gave a shorter proof of his theorem that the size of the shares in a quantum
secret sharing scheme must be as large as the secret itself.
Additionally, we proved that the size of a key in a
quantum Vernam cipher must be as large as the message itself.

It is an interesting open problem to prove better lower bounds on the
information rate for specific access structures different from
threshold schemes.

After this work was concluded, we were informed that the lower bound on the
size of shares of quantum secret sharing schemes has been proved in
\cite{Ogawa} by using different methods.

\section*{Acknowledgements}
Part of this research was funded by the project {\sc ProSecCo} under
IST-FET-39227.

\end{document}